\documentclass[a4paper,fleqn,usenatbib]{mnras}

\usepackage{newtxtext,newtxmath}

\usepackage[T1]{fontenc}
\usepackage{ae,aecompl}
\usepackage{pdflscape} 

\usepackage{graphicx}	
\usepackage{amsmath}	
\usepackage{amssymb}	

\title{A morphological study of galaxies in ZwCl0024+1652, a galaxy cluster at redshift z\,$\sim$\,0.4}
\author[Z. B. Amado et al.]{
Zeleke Beyoro Amado$^{1,\,2}$\thanks{E-mail: zbamado@gmail.com or zelekeb@essti.gov.et},
Mirjana Povi\'c$^{1,\,3}$,
Miguel S\'anchez-Portal$^{4,\,5,\,6}$, \newauthor 
S. B. Tessema$^{1}$,
\'Angel Bongiovanni$^{7,\,8,\,9}$,
Jordi Cepa$^{7,\,8,\,9}$,\newauthor
Miguel Cervi\~no$^{10}$,
J. Ignacio Gonz\'alez-Serrano$^{11}$, 
Jakub Nadolny$^{7,\,8}$, \newauthor
Ana Maria P\'erez Garcia$^{12}$,
Ricardo P\'erez-Martinez$^{13}$ and
Irene Pintos-Castro$^{14}$ \\
$^{1}$Ethiopian Space Science and Technology Institute (ESSTI), Entoto Observatory and Research centre (EORC),\\
\ \ Astronomy and Astrophysics Research and Development Division, P.O.Box 33679, Addis Ababa, Ethiopia\\
$^{2}$Kotebe Metropolitan University, College of Natural and Computational Sciences, Department of Physics, \\
\ \ P.O.Box 31248, Addis Ababa, Ethiopia\\
$^{3}$Instituto de Astrofisi\'ca de Andaluci\'a (IAA-CSIC),Glorieta de la Astronomia s/n, 18008, Granada, Spain\\
$^4$Instituto de Radioastronom\'ia Milim\'etrica, Av. Divina Pastora 7, N\'ucleo Central, E-18012 Granada, Spain\\
$^5$European Southern Observatory, Alonso de C\'ordova 3107, Vitacura, Santiago 763-0355, Chile\\
$^6$Joint ALMA Observatory, Alonso de C\'ordova, 3107, Vitacura, Santiago 763-0355, Chile\\
$^{7}$Instituto de Astrof\'isica de Canarias (IAC), E-38200 La Laguna, Tenerife, Spain\\
$^{8}$Departamento de Astrof\'isica, Universidad de La Laguna (ULL), E-38205 La Laguna, Tenerife, Spain\\
$^{9}$Asociaci\'on Astrof\'isica para la Promoci\'on de la Investigaci\'on, Instrumentaci\'on y su Desarrollo, ASPID, \\
\ \ E-38205 La Laguna, Tenerife, Spain\\
$^{10}$Centro de Astrobiolog\'{i}a (CSIC/INTA), 28850 Torrej\'{o}n de Ardoz, Madrid, Spain\\
$^{11}$Instituto de F\'{i}sica de Cantabria (CSIC - Universidad de Cantabria), Avda. de los Castros s/n, E-39005, Santander, Spain\\
$^{12}$Centro de Astrobiolog\'{i}a (CSIC/INTA), ESAC Campus, Camino Bajo del Castillo s/n, E-28692, Villanueva de la Ca\~nada, Spain\\
$^{13}$ISDEFE for ESA. Camino Bajo del Castillo s/n. Urb. Villafranca del Castillo. E-28692, Villanueva de la Ca\~nada, Spain\\
$^{14}$Department of Astronomy \& Astrophysics, University of Toronto, 50 St. George Street, Toronto, ON M5S 3H4, Canada}

\date{Accepted 2019 February 7. Received 2019 February 7; in original form 2018 November 25}

\pubyear{2019}

\begin{document}
\label{firstpage}
\pagerange{\pageref{firstpage}--\pageref{lastpage}}
\maketitle

\begin{abstract}
The well-known cluster of galaxies ZwCl0024+1652 at $z\sim0.4$, lacks an in-depth morphological classification of its central region. While previous studies provide a visual classification of a patched area, we used the public code called \underline{gal}axy \underline{S}upport \underline{V}ector \underline{M}achine (galSVM) and HST/ACS data as well 
as WFP2 master catalogue to automatically classify all cluster members up to 1\,Mpc. galSVM analyses galaxy morphologies through Support Vector Machine (SVM).
From the 231 cluster galaxies, we classified 97 as early-types (ET) and 83 as late-types (LT). The remaining 51 stayed unclassified (or undecided, UD). By cross-matching our results with the existing visual classification, we found an agreement of 81\%. In addition to previous Zwcl0024 morphological classifications, 121 of our galaxies were classified for the first time in this work. In addition, we tested the location of classified galaxies on the standard morphological diagrams, colour-colour and colour-magnitude diagrams. Out of all cluster members, $\sim$\,20\% are emission line galaxies (ELG), taking into account previous GLACE results.  
We have verified that the ET fraction is slightly higher near the cluster core and decreases with the clustercentric distance, while the opposite trend has been observed for LT galaxies. We found higher fraction of ET (54\,\%) than LT (46\,\%) throughout the analysed central region, as expected. In addition, we analysed the correlation between the five morphological parameters (Abraham concentration, Bershady-Concelice concentration, Asymmetry, Gini and M20 moment of light) and clustercentric distance, without finding a clear trend. Finally, as a result of our work, the morphological catalogue of 231 galaxies containing all the measured parameters and the final classification is available in the electronic form of this paper. 

\end{abstract}




\section{Introduction}
\label{sec:intro}

A consolidated observational fact is the outstanding difference in the properties of galaxies located in the cores (or regions of high local galaxy density) and in the external parts (or low density ones) of  low- and intermediate- redshift clusters: the former regions are dominated by red, massive and passive early-type galaxies (ET galaxies, comprising elliptical and S0), while a substantial increase of the fraction of late-type galaxies (LT, comprising spiral and irregular objects) is observed in the latter.  This was early identified by  \cite{Zwicky1942}, and quantified by \cite{Dressler1980} in the so-called morphology-density relation linking the increasing fraction of ET galaxies with local galaxy density.  Similarly, a decrease of the fraction of star forming (SF) galaxies is observed with increasing local galaxy density \citep[the SF-density relation, see for instance ][ and references therein]{PC2013}.  Moreover, these relations evolve with cosmic time, as was realized by \cite{ButOem1978}, who found that cluster galaxy populations evolve as redshift changes in such a way 
that rich clusters at higher redshift (z\,$>$\,0.2) are populated with a higher fraction of blue galaxies than low redshift clusters. This is the so-called Butcher-Oemler (BO) effect. Likewise, an increase of the cluster SF and active galactic nuclei (AGN) activity is observed \citep[see for instance ][]{Haines2009,Martini2013}.

The morphology-density relation seems to hold from nearby clusters up to redshifts as high as z\,$\sim$\,1.5 \citep[e.g. ][]{Dressler1997, Postman2005, Holden2007, Mei2012, Nantais2013}.  Likewise,  star formation takes place in low density regions where LTs dominate while 
high density regions are dominated by quiescent ET galaxies since $z\,\sim\,1.5$ to the local universe \citep[e.g. ][]{PosGe1984,Kauff2004,Cooper2012,Wetzel2012,Woo2013}. At higher redshift, there is some controversial evidence of the existence of a reversal of the SF-density relation: some authors, as \cite{Tran2010} find an (even dramatic) increase of the  fraction of SF galaxies from low- to high-density regions in clusters at z\,$\sim$\,1.6, while other authors \citep[e.g.][]{Ziparo2014}  do not find a clear evidence of such type of reversal when studying clusters at the same redshift.  \cite{Quadri2012}, using mass-selected samples from the UKIDSS Ultra-Deep Survey, conclude that galaxies with quenched SF tend to reside in dense environments out to at least z\,$\sim$\,1.8. 

The structural and morphological properties of a galaxy  are important tracers of its evolutionary stage. Thus, the correlation of the morphology (and/or SF activity) of the clusters' galaxies with the local density provides valuable information on the stage of infall at which galaxies experience the bulk of their transformations. To this end, it is important to perform wide-area surveys (to study the density-dependent effects) in clusters that span a range of redshifts (to assess the evolution with cosmic time). 

The morphological taxonomy of galaxies can be backdated to \citet{Reynolds1920}. Visual inspection is the traditional method, and even now a very common way to perform morphological classification of galaxies \citep[e.g.][]{Lintott2008,NA2010,Fasano2012,Kocevski2012,Kartaltepe2012,Kartaltepe2015,Buitrago2013,KumShamir2016,Willett2013,Simmons2017,Willett2017}. One of the drawbacks of the visual classification method is the subjectivity, that can be alleviated by performing multiple instances of the classification of each object carried out by different persons; an outstanding example is Galaxy Zoo project \citep[][]{Lintott2008,Lintott2011}. In the framework of this ``citizen science'' initiative, nearly one million galaxies from the Sloan Digital Sky Survey (SDSS) were classified by $\sim$\,10$^5$ participants who performed more than 4\,$\times$\,10$^7$ classifications. Needless to say, when dealing with large number of sources, the visual classification method can be really time-consuming. It works well for closer and well resolved objects for accurate estimation.  For such objects it agrees with the results of modern classification methods. 

But with a currently overgrowing observational astronomical data, the above method is probably not the most appropriate or even unfeasible for high-redshift galaxies. Modern classification techniques include galaxy fitting algorithms, which can give reliable results for a large number of galaxies in a relatively shorter period and with minimal human resources. 
To deal with fast growing and big astronomical data, machine learning techniques employing Convolutional Neural Networks (CNN) are widely under use recently for morphological classification of galaxies \citep[e.g. ][]{Banerji2010,Kuminski2014,DieWilDam2015,HC2015,AnThor2017,Domingues2018,Lukic2018}.
Modern galaxy classification methods can either be \textit{parametric} or \textit{non-parametric}. 

\textit{Parametric methods} use some parameters of the galaxies to classify them by fitting (one or two dimensional) mathematical models to their images assuming some predefined parametric model. 
In this approach S\'ersic profile \citep{Sersic1963} and a two-component profile (bulge + disk decomposition) are the commonly used models. The classification is obtained by fitting a two component profile as described in detail by \citet{Simard2002} and \citet{Peng2002}. 
More recently, \citet{Simard2011} has performed a classification of 1.12\,million galaxies using a bulge\,+\,disk decomposition approach with SDSS data release seven \citep{Abazajian2009}. In addition to this, a structural and morphological catalogue of 45\,million sources have been presented by \citet{Tarsitano2018}
with the Dark Energy Survey \citep{DES2016} data of the first year observation employing both a single S\'ersic parametric fits and non-parametric methods.
Parametric method in general is useful in that it gives a complete set of parameters describing the 
quantitative morphology. Since a large number of parameters need to be fitted, the results may be degenerated as shown in \citet{HC2007}.
Degeneracy occurs as a result of correlation between parameters, the results of the local minima in the parameter space of the chi-square minimization, or by numerical divergence in the process of fitting \citep{Peng2002,Peng2010}.
The peculiar characteristic of the parametric method in general is the assumption that a galaxy is described well by a simple analytic model whereas 
this does not always work for well resolved as well as irregular and merging/interacting galaxies.

On the other hand, the  \textit{non-parametric} approach  does not assume any specific analytic model and is performed on the basis of 
measuring a set of well-chosen observables. The effects of seeing, being one of the major challenges in galaxy fitting, are not included in non-parametric measurements unlike the parametric ones where the assumed mathematical model is convolved with the PSF. 
The non-parametric method was introduced for the first time by \citet{Abr1994,Abr1996} 
with the definition of two observables: the Abraham concentration index and asymmetry. A third quantity, namely smoothness, was introduced by \citet{Con2000,Con2003}. The classification has been further enhanced with additional observables: the GINI coefficient (\citealt{Abr2003}); 
M20 moment of light (\citealt{Lotz2004}) and Conselice-Bershady concentration (\citealt{Con2000, Bersh2000}). These six parameters, together with ellipticity are described in more details in subsection ~\ref{sec:Parameters}. 
\textit{Non-parametric} methods are in advantage when classifying large sample of galaxies at higher redshifts, when lower resolution data are available \citep[e.g. ][]{Scarlata2007,Tasca2009,Povic2009,Povic2013,Povic2015,PC2016}. Furthermore, no analytic predefined profile is required in this approach.\\ 
\indent In this paper, we apply a \textit{non-parametric} classification method to a well-known intermediate redshift cluster, namely ZwCl0024+1652 at z\,$=$\,0.395. This cluster has been extensively studied by several 
groups \citep[e.g. ][]{Morrison1997, Broadhurst2000, Kneib2003, Treu2003, Moran2007, Geach2009, Natarajan2009, Sanchez2015}. 
In particular, it has been observed by our team in the framework of the GaLAxy Cluster Evolution Survey \citep[GLACE; ][]{Sanchez2015} 
in the H$\alpha$ and [N{\sc ii}] emission lines to trace the SF and AGN activity in a wide range of environments (see Sect. \ref{sec:Discussions} below). 
Although this cluster has been deeply studied in some aspects, a visual morphological classifications has been performed for a 
limited number (214) of member galaxies within a clustercentric distance extending to 5\,Mpc \citep{Moran2007}. 
The purpose of this work is to improve the knowledge about the morphological properties of the member galaxies by providing 
a reliable classification of the sources up to 1 Mpc of clustercentric radius. We use publicly available Hubble Space Telescope (HST) 
Advance Camera for Survey (ACS) data in F775W filter.\\
\indent Only 66 galaxies have been classified by \citet{Moran2007} within this radius ($\simeq$31\,\% of the total sample). 
Second important objective of this work is to compare visual classification of \citet{Moran2007} with our \textit{non-parametric} 
method \-- this may provide us with handy-tool for future works at higher redshifts.\\  
\indent In our present work, we use a \textit{non-parametric} method called galSVM introduced by \citet{HC2008}. 
galSVM fits a number of parameters simultaneously and 
assigns probabilities for each galaxy to be classified. Then based on the probabilities, the galaxies are classified 
into two broad morphological classes, namely early-type (ET) and late-type (LT). For more details about this classification, we refer the reader to Section~\ref{sec:class}.\\
\indent The paper is organized as follows: Section~\ref{sec:Data} describes the data, along with a brief 
description of the generated source catalogue. In Section~\ref{sec:Morpho_class} the galSVM code and its 
application to our sample are described. The analysis on the results from our classification are further developed in Section~\ref{sec:Analysis}. 
A detailed discussion is presented in Section~\ref{sec:Discussions}.  Finally Section~\ref{sec:Summary} presents brief conclusions of this work.\\
\indent The following cosmological parameters are assumed throughout this paper: $\Omega_M=0.3$, $\Omega_\Lambda=0.7$, $\Omega_k=0$ 
and $H_0=70Kms^{-1}Mpc^{-1}$. All magnitudes are given in AB system as described by \citet{OkeGun1983}, unless otherwise stated.

\section{Data}
\label{sec:Data}
%
\subsection{HST/ACS data}
\label{sec:ACS}
\begin{figure*}
    \includegraphics[width=17cm]{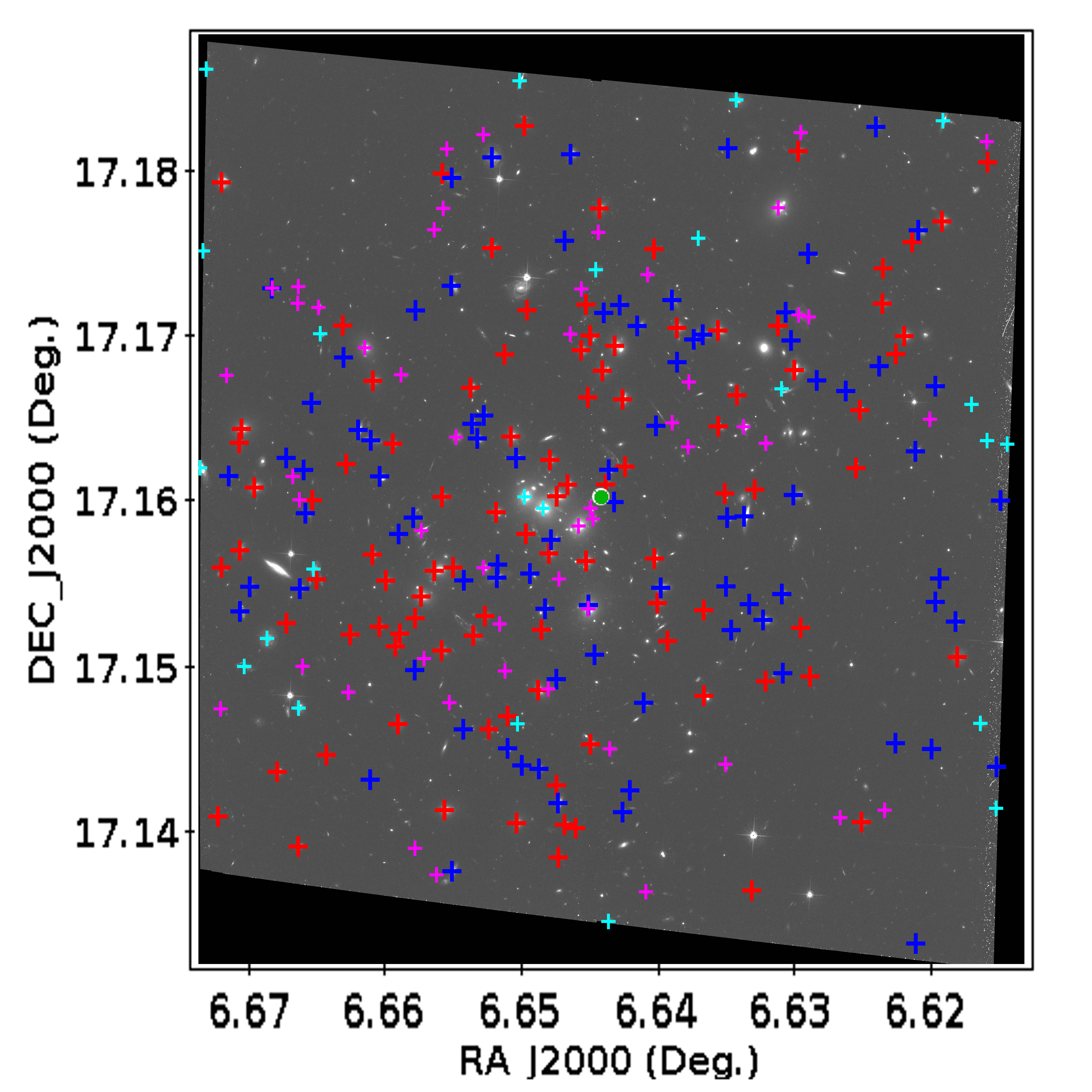}
    \caption{The ACS/WFC image of ZwCl0024+1652 galaxy cluster used in this work, where east is to the left and north at the top. The centre of the cluster is indicated by a large green dot. The larger red and blue crosses indicate galaxies classified as ET and LT, respectively. While the smaller cyan crosses show those galaxies for which the probabilities are not measured 
    and the magenta crosses show galaxies with measured probabilities but undecided morphologies (see Sec.~\ref{sec:Morpho_class}).}
    \label{Fig:ACS_image}
\end{figure*}

We used the public HST reduced scientific image of ZwCl0024+1652\footnote{Based on observations made with 
the NASA/ESA HST, and obtained from the Hubble Legacy Archive, which is a collaboration between the Space Telescope Science Institute (STScI/NASA), the Space Telescope European Coordinating Facility (ST-ECF/ESA) and the Canadian Astronomy Data centre (CADC/NRC/CSA)} from the observation made on 16 November 2004 with the ACS Wide Field Camera (WFC) using the F775W filter. The ACS/WFC has a pixel scale of $0.05\ arcsec/pixel$ and field of view of $202\times202\ arcsec^2$. The cluster is centred at RA = $6.64433\deg$ and DEC = $17.16211\deg$, and the used image covers the central part of cluster of $\sim$\,1\,Mpc. The image data is shown in Fig.~\ref{Fig:ACS_image} with all the sources labeled. 
\subsection{WFP2 supercatalogue data}
\label{sec:WFP2}
\begin{figure}
    \includegraphics[width=\columnwidth]{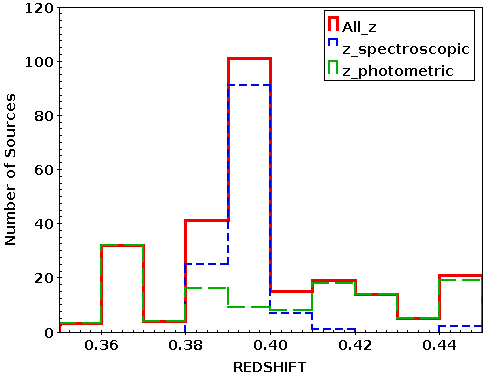}
    \caption{Redshift distribution of our real ZwCl0024+1652 sample where the red solid lines stands for the total sample, the blue dashed lines for spectroscopic redshifts and the dashed green lines for photometric redshifts.}
    \label{Fig:z_dist}
\end{figure}

To extract redshift information and to identify cluster members, we used the public ZwCl0024+1652 master catalogue\footnote{http://www.astro.caltech.edu/$\sim$smm/clusters/} described in \citet{Treu2003} and 
\citet{Moran2005}. The catalogue consists of 73,318 sources, with photometric and/or spectroscopically confirmed redshifts available, and covering the area of 0.5\,$\times$\,0.5\,deg$^2$ up to the clustercentric distance of about 5\,Mpc. All observations were carried out with the Canada-France-Hawaii Telescope (CFHT) and its CFH12K wide field camera, and/or the HST Wide Field and Planetary Camera (WFP2), as described in \citet{Treu2003}. Beside redshifts, this catalogue includes the visual morphological classification of sources brighter than I\,=\,22.5 \citep{Moran2005}. We cross-matched this catalogue with our SExtractor catalogue (3515 sources) using a maximum radius of 2\,arcsec. This radius was selected after testing different ones from 1 to 5\,arcsec and finding it to be the best compromise between being the counterparts and having multiple matches. 
We obtained a total of 255 counterparts (hereafter cluster sample) with available redshifts. In total, 126 and 129 sources have spectroscopic and photometric redshift measurements. The redshift distribution of the members is given in Fig.~\ref{Fig:z_dist}, including spectroscopic and photometric measurements.  
\section{Morphological Classification}
\label{sec:Morpho_class}
In this section we describe the morphological classification of the ZwCl0024+1652 cluster galaxies in detail. We first go briefly through the methodology used, obtained results, and final classification.   
 \subsection{Methodology}
 \label{sec:galSVM}
In this work we use galSVM \citep{HC2008} to classify galaxies morphologically in the ZwCl0024+1652 cluster. The galSVM is a public code that uses a free library libSVM \citep{CL2011} and works in IDL environment. It has been successfully tested previously, at different redshifts, and on both field and cluster galaxies \citep[e.g.,][]{HC2009,HC2010,HC2011,Povic2012,Povic2013,Povic2015,PC2016}. 

For source detection, flux extraction, and measurement of the morphological parameters we need for the morphological classification (e.g., ellipticity), we run SExtractor \citep{BerArn1996}. We extracted 3515 possible sources, including cluster members and field galaxies. galSVM uses local sample with known visual morphologies (see Sec.~\ref{sec:Local}) and use it to learn how these may be seen at redshfits and magnitudes distributions of our real sample. It consists of several steps. First, it simulates local galaxies, by placing them to the redshift and magnitude distributions characteristic of the real sample. Secondly, it drops simulated local galaxies into the background that corresponds to the real sample image. Third, it measures different morphological parameters (see Sec.~\ref{sec:Parameters}), first of the simulated local sample, and then of the real sample of galaxies that we want to classify. Finally, it compares morphological parameters of the training simulated local galaxies with their known visual classification, and determines conditions inside the multiple-parameters space that are then applied to the real sample to be classified. The final classification is based on a number of Montecarlo (MC) simulations, where each simulation gives a probability that the galaxy is early-type (ET). The average probability ($P\_avg$) and measured error give the final classification that the galaxy is ET. The probability that galaxy is late-type (LT) will then be $1-P\_avg$. For more details regarding galSVM see \cite{HC2008}. For applying galSVM, to include morphology determination for more fainter galaxies we took a magnitude limit of F775W\,$\le$\,26.  
\subsection{Training sample of local galaxies}
 \label{sec:Local}
\begin{figure*}
    \includegraphics[width=15cm]{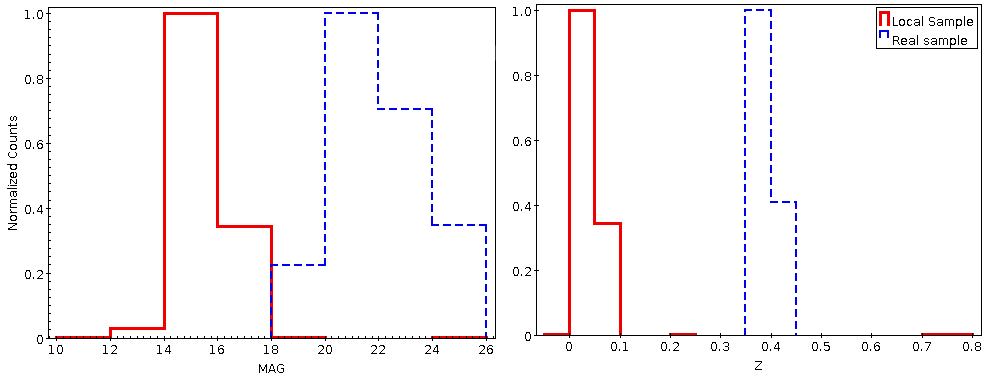}
    \caption{Magnitude \textit{(right plot)} and redshift \textit{(left plot)} distributions of the local training sample with known morphology (red solid lines) and the real ZwCl0024+1652 sample that should be classified (blue dashed lines).}
    \label{Fig:MAGZ_LOCAL}
\end{figure*}

We used a catalogue of 3000 visually classified local galaxies, with known 
redshifts and magnitudes. The sample was selected randomly from the \citet{NA2010} catalogue of visual morphology consisting of about 14000 galaxies taken from the Sloan Digital Sky Survey (SDSS)\footnote{SDSS is managed by the Astrophysical Research Consortium for the Participating Institutions of the SDSS Collaboration including the University of Arizona, the Brazilian Participation Group, Brookhaven National Laboratory, Carnegie Mellon University, University of Florida, the French Participation Group, the German Participation Group, Harvard University, the Instituto de Astrofisica de Canarias, the Michigan State/Notre Dame/JINA Participation Group, Johns Hopkins University, Lawrence Berkeley National Laboratory, Max Planck Institute for Astrophysics, Max Planck Institute for Extraterrestrial Physics, New Mexico State University, New York University, Ohio State University, Pennsylvania State University, University of Portsmouth, Princeton University, the Spanish Participation Group, University of Tokyo, University of Utah, Vanderbilt University, University of Virginia, University of Washington, and Yale University.} DR4 data. The redshift distribution of our local sample is in a range of 0.01\,-\,0.1, and most of galaxies are bright with r band magnitude between 13 and 17. The magnitude and redshift distributions of the training sample versus that of the real data are shown in the two plots of Fig.~\ref{Fig:MAGZ_LOCAL}. The detailed description of the training sample can be found in \citet{Povic2013}. The training sample of 3000 local galaxies was selected as a good compromise between the computing time and accuracy in classification (both being highly sensitive to the training sample size). In addition, equal number of ET and LT galaxies were taken into account to obtain more precise morphology, and the selected sample can be considered as representative of the whole data with respect to general galaxy properties, as shown in \citet{Povic2013}. 
  \subsection{Measured morphological parameters}
 \label{sec:Parameters}
We used the following six parameters simultaneously to run galSVM: ellipticity (obtained by SExtractor), asymmetry, Abraham concentration index, GINI coefficient, M20 moment of light, and Conselice-Bershady concentration index. The last five were measured using galSVM and are briefly described as follows.
\begin{enumerate}
\item \textbf{Asymmetry (ASYM),} measures an extent to which galaxy's light is rotationally symmetric \citep{Abr1994,Abr1996,Con2003}. 
\item \textbf{Abraham concentration index (CABR),} is defined as the ratio of fluxes of the inner isophote at 30\,\% to that of the outer isophote at 90\,\% \citep{Abr1994,Abr1996}. 
\item \textbf{GINI coefficient (GINI),} is a statistical term derived from the Lorentz curve specifying the overall distribution function of the pixel values of the galaxy \citep{Abr2003,Lotz2004}. 
\item \textbf{M20 Moment of light,} describes the second order normalized moment of the 20\,\% brightest pixels of the particular galaxy \citep{Abr2003,Lotz2004}.
\item \textbf{Bershady-Conselice concentration index (CCON),} measures a light ratio within a circular inner aperture (radii comprising of 20\,\% of the total flux) to the outer aperture (radii containing 80\,\% of the total flux) of the galaxy \citep{Con2000,Bersh2000,Con2003}. 
\end{enumerate}
In all measurements, the total flux is defined as the amount of flux contained within 1.5 times the Petrosian radius, where the Petrosian radius was measured with SExtractor. The centre of the galaxy is defined by minimizing ASYM index. More details on all parameters can be found in \cite{HC2008} and \cite{Povic2013}.  
\subsection{galSVM applied to ZwCl0024+1652}
 \label{sec:Applying_galSVM}
To measure the morphologies of ZwCl0024+1652 cluster members, we run galSVM on the HST/ACS F775W image described in Sec.~\ref{sec:ACS}, and used the SExtractor catalogue of 255 sources with all needed input parameters and redshifts available (see Sec.~\ref{sec:ACS} and \ref{sec:WFP2}). We went through all galSVM steps described in Sec.~\ref{sec:galSVM}, using the 3000 SDSS local galaxies as a training sample (see Sec.~\ref{sec:Local}). We measured all parameters described in Sec.~\ref{sec:Parameters} of both training and real samples. For final classification we run 15 MC simulations, where in each simulation we used 2000 different randomly selected local galaxies (out of 3000) with the same number of ETs and LTs. The number of MC simulations was selected as the best compromise between the computational time and accuracy of results (see \citealt{Povic2013}).

\begin{figure*}
    \includegraphics[width=15cm]{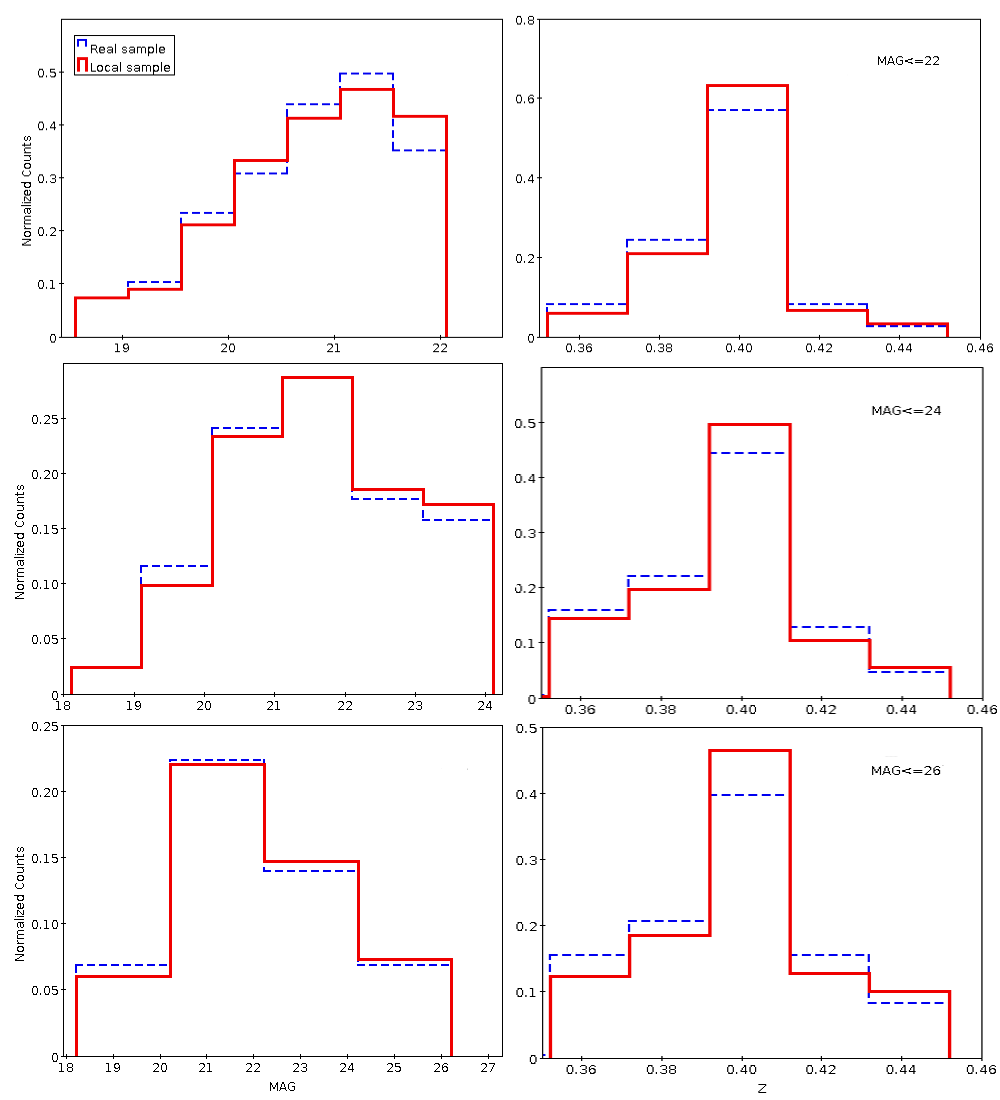}
     \caption{Magnitude \textit{(left plots)} and redshift \textit{(right plots)} distributions of our real data (blue dashed lines) and the simulated local sample (red solid lines). The distributions 
     are plotted for F775W\,$\leq$\,22.0 (top plots), F775W\,$\leq$\,24.0 (middle plots), and F775W\,$\leq$\,26.0 (bottom plots).}
     \label{fig:MAGZ}
\end{figure*}

Taking into account previous results obtained in \cite{Povic2013}, dividing a sample into different magnitude ranges can increase the accuracy of morphological classification by optimizing the galSVM code for fainter galaxies. Therefore in this work we run galSVM three times, using the following ranges.
\begin{enumerate}
  \item F775W\,$\leq$\,22.0 (137 galaxies),
  \item F775W\,$\leq$\,24.0 (216 galaxies), and
  \item F775W\,$\leq$\,26.0 (255 galaxies).
 \end{enumerate}
For each range we provided the corresponding magnitude and redshift distributions of cluster members for simulating during the classification process. These distributions are shown in Fig.~\ref{fig:MAGZ} for both training sample after being simulated and the real sample to be classified.   
For the final classification we follow the findings of \cite{Povic2013}, and considered the results from the first magnitude bin, from the second, but only for those sources not present in the first one (for 22.0\,$<$\,MAG\_AUTO\,$\leq$\,24.0), and finally from the third bin, but only for those sources not present in the previous two (for 24.0\,$<$\,MAG\_AUTO\,$\leq$\,26.0).
\subsection{Final classification}
\label{sec:class}
In all galSVM runs (i.e. for each magnitude bin) we obtain a final average probability from 15\,MC simulations (for more details about the training sample and the running setup see Sections~\ref{sec:Local} and ~\ref{sec:Applying_galSVM}). Finally, we obtained PROBA\_AVG\footnote{PROBA\_AVG is average probability measured by galSVM} with corresponding uncertainty values for 231 galaxies out of 255. For the remaining 24 galaxies, PROBA\_AVG was not measured either because one or more parameters have values out of the respective standard range \citep{HC2008}, or they simply were not measured by galSVM. Of these, 50\,\% are located on image borders, while most of the remaining sources are merging/interacting systems or some are edge on galaxies. Only 3 galaxies (out of 24) have close companions, but the sample is not statistically significant for doing any additional studies. Tables~\ref{tab:proba_st} and \ref{tab:error_st} give the median values and Q1 to Q3 ranges\footnote{Q1 stands for quartile1 (25\% of the sample) and Q3 for quartile3 (75\% of the sample)} of average probability [Q1\,-\,Q3]\footnote{[Q1\,-\,Q3] stands for the range of values characteristic of 50\% of the analysed sample} and its error in three magnitude bins. It can be seen that most of the brightest galaxies (F775W\,$\leq$\,22.0) are characterized by larger probability (median value > 0.7) to be ETs, while for fainter galaxies (2nd and 3rd bin) probability to be ET is lower than 0.5. As we could expect, the error values increase for fainter bins (see Table~\ref{tab:error_st}).   
\begin{table}
	\centering
	\caption{The PROBA\_FINAL values in the three magnitude ranges}
	\label{tab:proba_st}
	\begin{tabular}{lll}
		\hline\hline
		& Median &\,[Q1\,-\,Q3] \\
		\hline\hline
		F775W\,$\leq$\,22.0 & 0.745 & 0.263 - 0.877 \\
		22.0\,$<$\,F775W\,$\leq$\,24.0 & 0.474 & 0.170 - 0.727 \\
		24.0\,$<$\,F775W\,$\leq$\,26.0 & 0.478 & 0.271 - 0.569 \\
		\hline
	\end{tabular}
	\end{table}

\begin{figure}
    \includegraphics[width=\columnwidth]{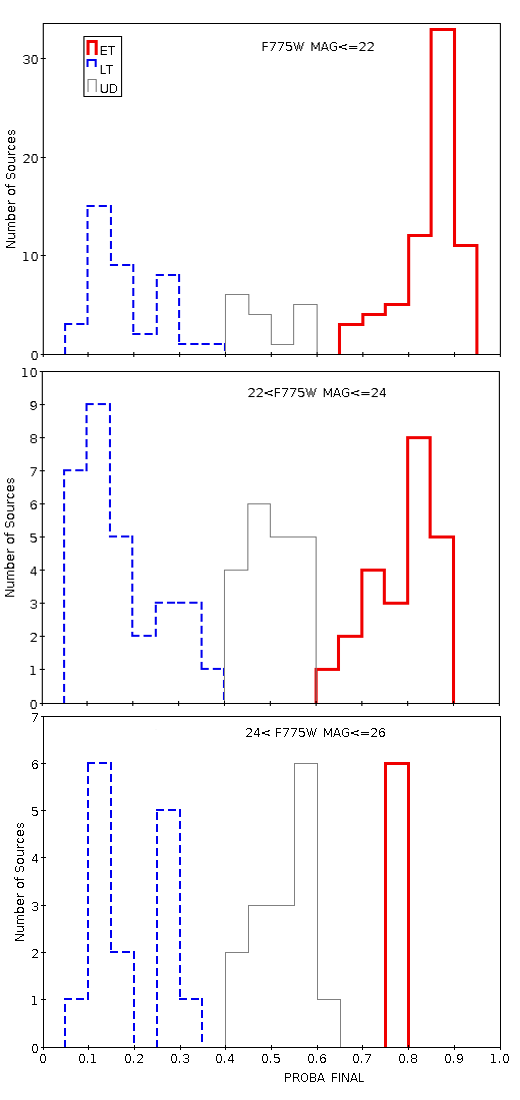}
    \caption{Final error corrected probability distributions of all galaxies classified as ET (red thick solid lines), LT (dashed blue lines), and UD (grey thin solid lines) in the three magnitude ranges.}
    \label{fig:Proba_merged}
 \end{figure}

\begin{figure}
    \includegraphics[width=8cm]{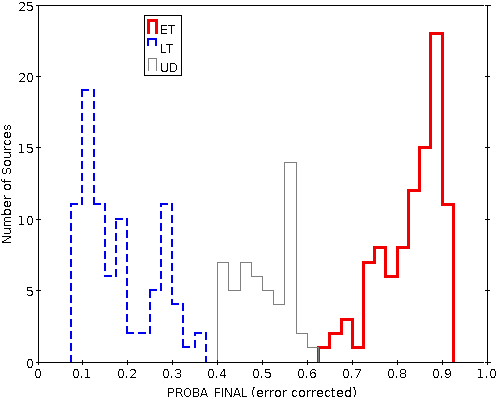}
    \caption{Final error corrected probability distributions of all ET (red thick solid lines), LT (dashed blue lines), and UD (grey thin solid lines}
    \label{fig:classified}
\end{figure}

\begin{table}
	\centering
	\caption{The PROBA\_ERR values in the three magnitude ranges}
	\label{tab:error_st}
	\begin{tabular}{lll} 
		\hline\hline
		& Median & [Q1 - Q3] \\
		\hline\hline
		F775W\,$\leq$\,22.0 & 0.044 & 0.030 - 0.067 \\
		22.0\,$<$\,F775W\,$\leq$\,24.0 & 0.045 & 0.032 - 0.059 \\
		24.0\,$<$\,F775W\,$\leq$\,26.0 & 0.069 & 0.054 - 0.122 \\
		\hline
	\end{tabular}
	\end{table}
For the final classification we took into account the measured errors and considered a galaxy to be ET if PROBA\_FINAL\,=\,PROBA\_AVG\,$\pm$\,PROBA\_ERR\,$>$\,0.6 (or 0.7 in the last magnitude bin), and to be LT if PROBA\_FINAL\,$<$\,0.4 (or 0.35 in the last magnitude bin), where PROBA\_ERR is uncertainty in measuring probability and PROBA\_FINAL is the final probability after error correction. For those galaxies with 0.4$<$\,PROBA\_FINAL\,$<$\,0.6 (or between 0.35 and 0.7 in the last magnitude bin) we are not able to classify them morphologically, and they will remain inside the 'undecided class' (UD). To define the classification boundaries, we used previous works of \cite{Povic2012,Povic2013} and \cite{PC2016}. Fig.~\ref{fig:Proba_merged} shows the PROBA\_FINAL distributions in the three magnitude ranges, while the final classification is summarized in Table~\ref{tab:class_all}. As can be seen, out of a total of 231 galaxies with measured final probabilities, we have 97 (42\%), 83 (36\%), and 51 (22\%) galaxies classified as ET, LT, and UD, respectively. Figure~\ref{fig:classified} shows the PROBA\_FINAL of the whole classified sample. 
We marked in Fig.~\ref{Fig:ACS_image} all the classified sources with red and blue crosses respectively for ET and LT galaxies respectively. Of the classified sources ET (LT) galaxies 59 (41) have spectroscopically confirmed and 38 (42) have photometric redshifts.
Few bright galaxies (e.g. close to the cluster centre) remained unclassified, mainly due to the difficulty that galSVM had with classifying galaxies being in rich environments and with close companions. Moreover, it has also been deduced that only 21.6\,\% of the UD galaxies have spectroscopically confirmed redshifts while for the classified ones (ET or LT), 62.2\,\% have spectroscopic redshifts.

\begin{table}
	\centering
	\caption{Final classification of the ZwCl0024+1652 cluster members.}
	\label{tab:class_all}
	\begin{tabular}{lcccc} 
			\hline\hline
		& \textbf{ET} & \textbf{LT} & \textbf{UD} & \textbf{Total} \\
		\hline\hline
		F775W\,$\leq$\,22.0 & 68 (55\%) & 39 (32\%)& 16 (13\%)& 123 \\
		22.0\,$<$\,F775W\,$\leq$\,24.0 & 23 (32\%)& 29 (40\%)& 20 (28\%)& 72 \\
		24.0\,$<$\,F775W\,$\leq$\,26.0 & 6 (17\%)& 15 (42\%)& 15 (41\%)& 36 \\
		\hline
		\textbf{Total} & \textbf{97} & \textbf{83} & \textbf{51} & \textbf{231} \\ 
		\hline
	\end{tabular}
\end{table}

\section{Analysis}
\label{sec:Analysis}
\subsection{Comparisons with visual morphological classification}
\label{sec:Xmorpho}
\begin{figure}
 \includegraphics[width=8cm]{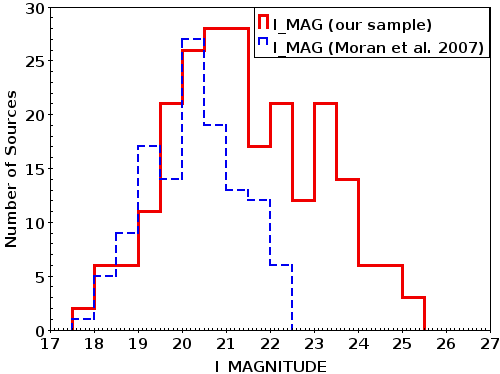}
\caption{Comparison between the I band magnitude (red solid line) of our sample galaxies classified in this work using galSVM and the I band magnitude of the counterparts from \citet{Moran2007} visually classified cluster members (blue dashed line).}
    \label{fig:Mag_compare}
 \end{figure}

The visual morphological classification of 214 galaxies with spectroscopically confirmed redshifts in ZwCl0024+1652 was carried out previously by \cite{Moran2007}, covering the clustercentric distance of 5\,Mpc. In this section we compare our non-parametric classification of 231 galaxies, within the clustercentric distance of 1\,Mpc (see Sec.~\ref{sec:Data}), with the visual one. 
Within the region of our data ($\sim$1\,Mpc radius) we determined that there are 123 sources with spectroscopically confirmed redshifts having visual morphology as in \citet{Moran2007}. While in our sample catalogue with measured probabilities (231 sources) we have 111 sources with spectroscopic redshifts and 120 sources with photometric redshifts. We cross-matched the two catalogues using the radius of 2\,arcsec, and found a total 66 counterparts. The reason of a small number of counterparts is mainly because of the fact that \cite{Moran2007} visual classification was done only for galaxies with confirmed spectroscopic redshifts and only considers the best resolved galaxies. The I-band magnitude limit of galaxies in \cite{Moran2007} is 22.3, with 201 (95\,\%) of galaxies being brighter than I\,=\,22, whereas our magnitude limit in F775W-band is 26. The I\_band magnitude distribution comparison of both works is given in Fig.~\ref{fig:Mag_compare}.
\indent Out of 66 counterparts, 50 and 16 galaxies were classified visually by \cite{Moran2007} as ET and LT respectively. When compared with our results, 53 galaxies, or 81\,\%, match the visual classification, of these 41 being classified as ET, and 12 as LT. Of the remaining 13 galaxies, 7 have visual classification available, but were classified as UD in our work, while for the other 6 galaxies ET/LT classification is in disagreement between the two works. 
Visually checking these galaxies, we found that 3 of them are edge on (S0 in \citet{Moran2007} while LT in our work possibly Sa that both could be possible). The other three galaxies were classified as ET in in our work whereas 2 of them are Sa\,+\,b and one is Sc\,+\,d in \citet{Moran2007}; our classification being right for one while one is observed to be an interacting system and the remaining one is peculiar galaxy.
Finally, after these comparisons we conclude that 81\,\% of our classification is in a good agreement with the visual classification. Moreover, in this work we provide a reliable classification of additional 121 galaxies within 1\,Mpc of clustercentric distance, being classified for the first time.   
\subsection{Morphological parameters}
\label{sec:histo}
\begin{figure*}
    \includegraphics[width=15cm]{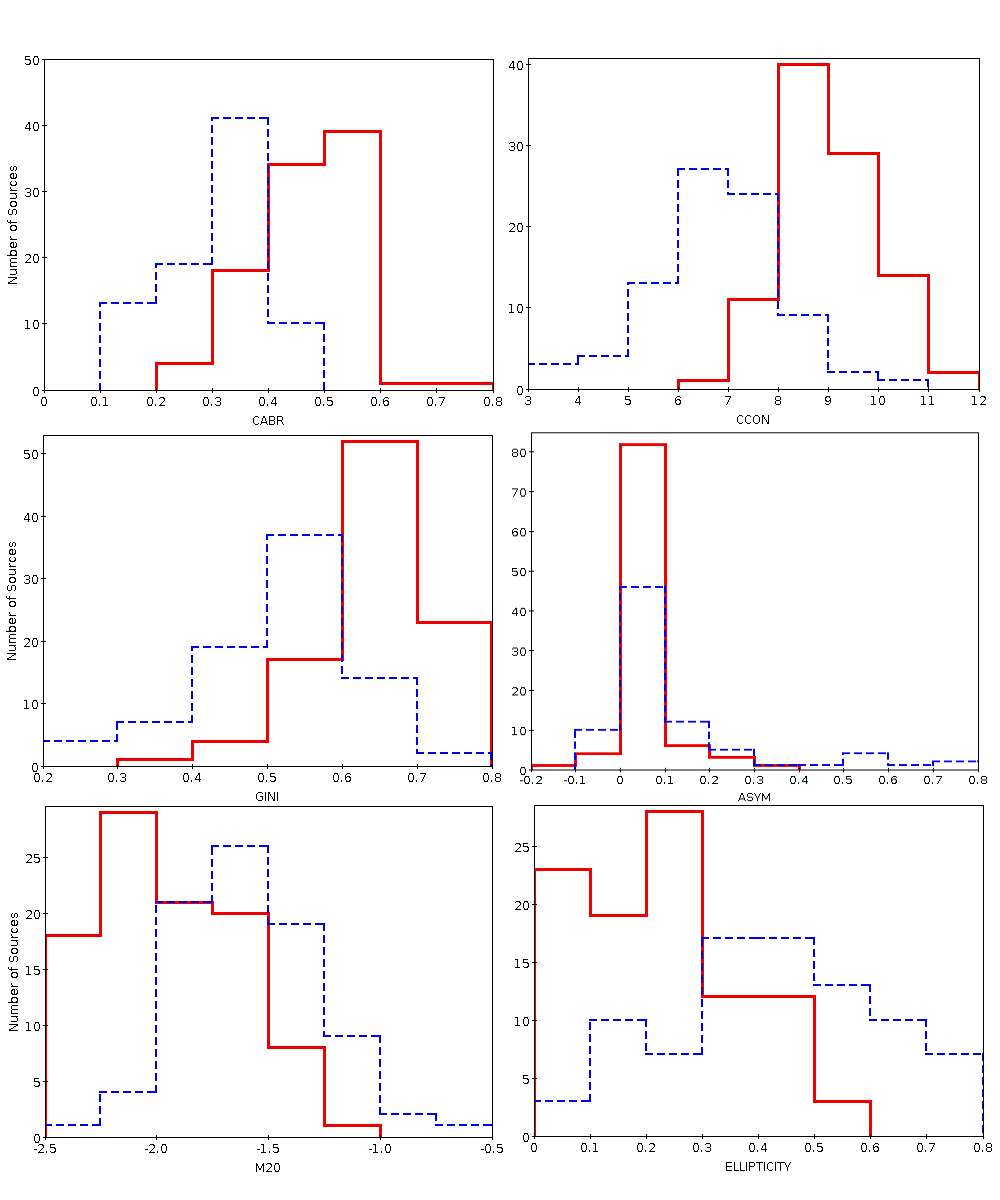}
    \caption{\textit{(From top left to bottom right:)} Distributions of CABR, CCON, GINI, ASYM, M20 moment of light, and ellipticity parameters of ET (red solid lines) and LT (blue dashed lines) galaxies.}
    \label{fig:Para_histo}
\end{figure*}

The distributions of different measured morphological parameters of 180 ET and LT classified cluster members are given in Fig.~\ref{fig:Para_histo}. In addition to the histograms, Table~\ref{tab:Para_table} summarizes the median values of each parameter and [Q1-Q3] range characteristic of cluster members classified as ET or LT. As can be seen from both Fig.~\ref{fig:Para_histo} and Table~\ref{tab:Para_table}, all parameters follow the expected trends of ET and LT galaxies, with concentration indices such as CABR, CCON and GINI being characterised with higher values in case of ETs, while ASYM, M20, and ELLIP are showing higher values for LTs. If we compare our results with those obtained by \cite{Povic2013}, using the same methodology and data of the ALHAMBRA survey \citep{Moles2008} in F613W band, ZwCl0024+1652 galaxies classified as ET seem to be slightly more concentrated (in terms of all concentration indices), and characterised with lower asymmetries in the case of both ET and LT.    

\begin{table}
	\centering
	\caption{Median and [Q1\,-\,Q3] range of measured morphological parameters of galaxies classified as ET or LT.}
	\label{tab:Para_table}
	\begin{tabular}{llll} 
	
		\hline\hline
		\textbf{Parameter} & \textbf{Measure} & \textbf{ET} & \textbf{LT} \\
		\hline
		CABR & median & 0.493 & 0.322 \\
		     & [Q1\,-\,Q3] & 0.430 - 0.532 & 0.268 - 0.374 \\
		\hline
		CCON & median & 8.894 & 6.865 \\
		     & [Q1\,-\,Q3] & 8.265 - 9.548 & 6.020 - 7.470 \\
		\hline
		GINI & median & 0.650 & 0.534 \\
		     & [Q1\,-\,Q3] & 0.614 - 0.694 & 0.452 - 0.587 \\
		\hline
		ASYM & median & 0.040 & 0.062 \\
		     & [Q1\,-\,Q3] & 0.025 - 0.059 & 0.021 - 0.158 \\
		\hline
		M20 & median & -1.996 & -1.597 \\
		     & [Q1\,-\,Q3] & -2.195 -- -1.726 & -1.805 -- -1.401 \\
		\hline
		ELLIP & median & 0.224 & 0.433 \\
		     & [Q1\,-\,Q3] & 0.102 - 0.307 & 0.303 - 0.585 \\
		     \hline
		\end{tabular}
		\end{table}

\subsection{Morphological diagnostic diagrams}
\label{sec:diagrams}
\begin{figure*}
    \includegraphics[width=17cm]{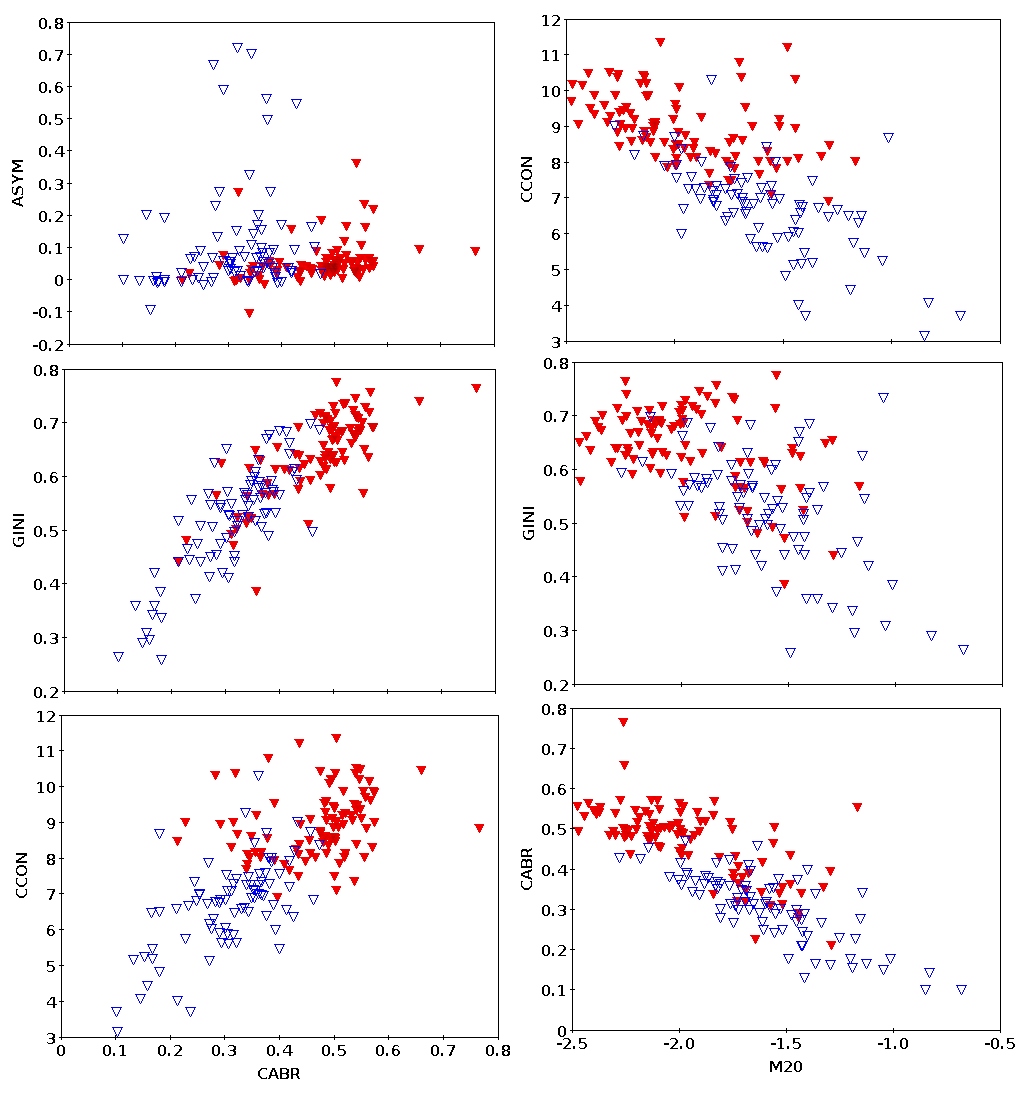}
    \caption{Standard morphological diagnostic diagrams showing the relation between CABR and ASYM, GINI, and CCON (left plots, from top to bottom, respectively), and M20 and CCON, GINI, and CABR (right plots, from top to bottom, respectively). In all the plots red solid and blue open triangles stand for ET and LT galaxies, respectively.}
    \label{fig:SR}
\end{figure*}

In this section, we tested some of the commonly used morphological diagnostic diagrams by comparing the measured morphological parameters. Fig.~\ref{fig:SR} shows six different diagrams and relations between CABR and ASYM, GINI, and CCON (left plots, from top to bottom, respectively), and M20 and CCON, GINI, and CABR (right plots, from top to bottom, respectively). These relations have been used in many previous works, showing a separation between ET and LT galaxies \citep[e.g.,][]{Abr1994,Abr1996,Con2000,Abr2003,Con2003, Lotz2004,Cassata2007,Scarlata2007,Tasca2009,Povic2009,Povic2013,PC2016,Tarsitano2018}. 

It can be seen in all plots that ZwCl0024+1652 cluster members classified as ET and LT are occupying different areas on diagrams, as expected. ETs are located again in the regions characterised with higher concentrations (larger values of CABR, GINI, and CCON and lower of M20), in comparison to LTs. ASYM parameter is much more delicate in separating sources, as has been commented previously \citep{Povic2015}, and as can be seen in Fig.~\ref{fig:SR} (top left plot). However, it can be efficient in selecting interacting systems, showing larger values as can be seen in the same plot for sources with ASYM\,$>$\,0.5. 
The relationships depicted are all in agreement with recent works (e.g. \citealt{Cast2014}; \citealt{Parekh2015}; \citealt{PC2016}).
We took into account studies of \citet{Tarsitano2018} and their Figure 11. We reproduced Gini vs. M20 diagram color coded with CABR, CCON, ASYM, and ELLIP finding the results in line with our Fig.~\ref{fig:SR}.

\subsection{Colour-colour and colour-magnitude relations}
\label{sec:CC}
\begin{figure*}
    \includegraphics[width=18cm]{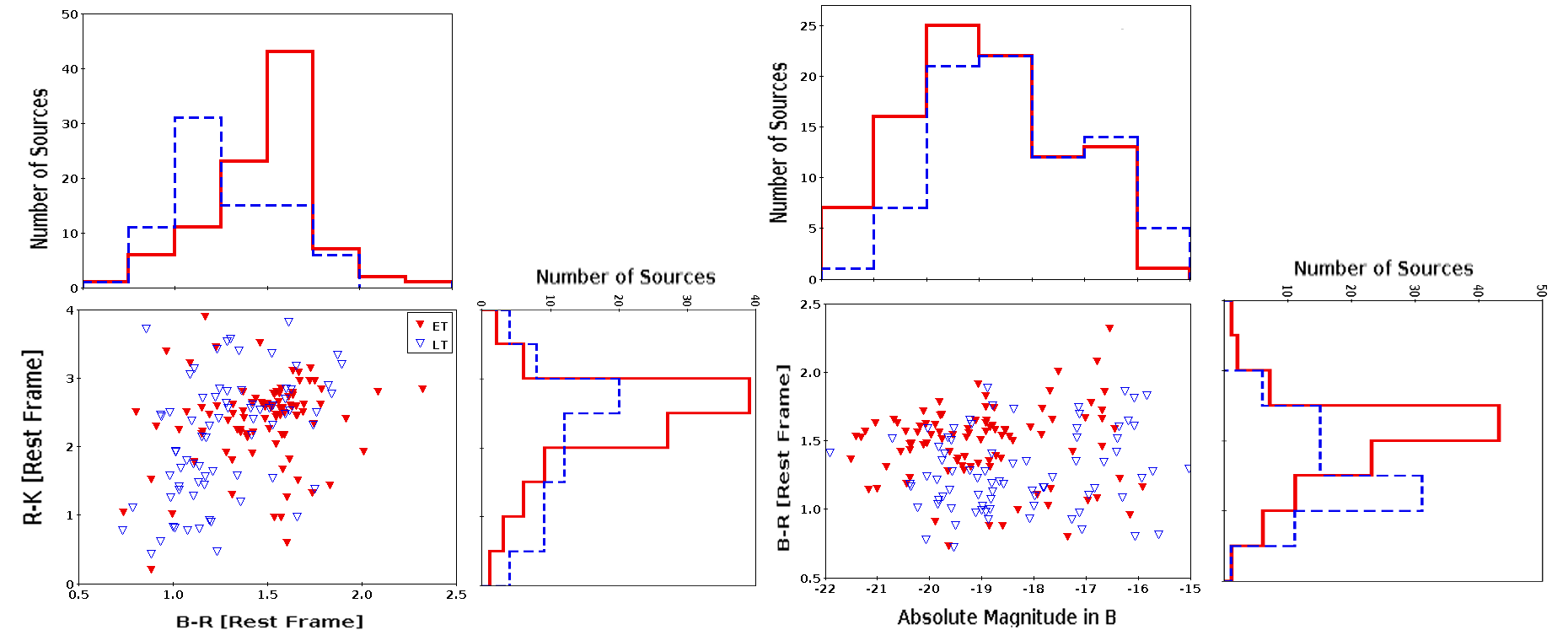}
    \caption{Rest frame R\,-\,K vs. B\,-\,R colour-colour diagram (left plot), and 
    B\,-\,R rest-frame colour and absolute magnitude in B diagram (right plot)}
    \label{fig:CMD}
\end{figure*}

In this section we tested the colour-colour and colour-magnitude diagrams for ZwCl0024+1652 cluster members classified as ETs and LTs. These diagrams have been tested at both lower and higher redshifts, and it is very well known that the distribution of galaxies on them is bimodal, with ETs being mainly located in the red sequence and LTs in the blue cloud \citep[e.g.,][etc.]{Bell2003,Cassata2007, Melbourne2007,Povic2013,Schawinski2014}. In Fig.~\ref{fig:CMD} we represented the relation between the R\,-\,K versus B\,-\,R rest-frame colours (left plot), and between the B\,-\,R rest-frame colour and absolute magnitude in the B band (right plot). We also represent the histograms of all parameters used in the 2d plots, and their distributions for both ET and LT galaxies. In the two plots we can see the area with a higher density of ET sources, and that in general brighter and redder regions have higher fractions of ETs, as expected, while fainter and bluer parts of the diagram are populated more with LTs. 
\subsection{Morphology vs. clustercentric distance}
\label{sec:D_gc}
\begin{figure}
    \includegraphics[width=\columnwidth]{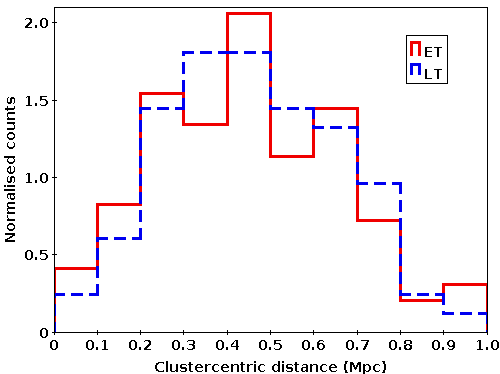}
    \caption{Normalised distribution of the clustercentric distance of galaxies classified as ET (red solid lines) and LT (blue dashed lines).}
    \label{fig:D_histo}
\end{figure}

The distance between member galaxy and the centre of the cluster is calculated using the spherical law of cosines as: 
\begin{equation}
cos(D_s) = sin(\delta_c) \times sin(\delta_g) + cos(\delta_c) \times cos(\delta_g) \times cos(|\alpha_c - \alpha_g|),
 \label{eq:D1}
\end{equation}
where $(\alpha_c,\delta_c)$ are right ascension and declination of the cluster centre in radians, while $(\alpha_g,\delta_g)$ are galaxy coordinates. To measure the clustercentric distance in Mpc, we used the following:
\begin{equation}
R = D_{cl} \times tan(D_s)\simeq D_{cl}\times D_s,
 \label{eq:D4}
\end{equation}
where $D_{cl}=1500Mpc$ and is the distance to ZwCl0024+1652. Fig.~\ref{fig:D_histo} shows the distribution of clustercentric distance of 180 cluster members classified as ETs and LTs, while Table ~\ref{tab:D} provides the basic statistics (median and [Q1\,-\,Q3] ranges) of both morphological types. 

\begin{table}
	\centering
	\caption{Statistical analysis of the morphological class distribution with respect to the clustercentric 
	distance $(R)$, the MUMEAN and the F775W\_MAG values. Here the median and the value range $[Q1 - Q3]$ for 50\% 
	of the sources in each class to fall is computed}.
	\label{tab:D}
		\begin{tabular}{llll} 
		\hline\hline
		\textbf{Parameter} & \textbf{Measure} & \textbf{ET} & \textbf{LT} \\
		\hline
		$R$ & median & 0.455 & 0.477 \\
		     & $[Q1 - Q3]$ & 0.308 - 0.623 & 0.324 - 0.636 \\
		     \hline
	        MUMEAN & median & 22.022 & 22.120 \\
		     & $[Q1 - Q3]$ & 21.925 - 22.105 & 22.074 - 22.143 \\
		     \hline
		F775W\_MAG & median & 21.284 & 22.169 \\
		     & $[Q1 - Q3]$ & 20.324 - 22.538  & 21.384 - 23.492 \\ 
		\hline
		\end{tabular}
		\end{table}
		
\begin{figure}
    \includegraphics[width=\columnwidth]{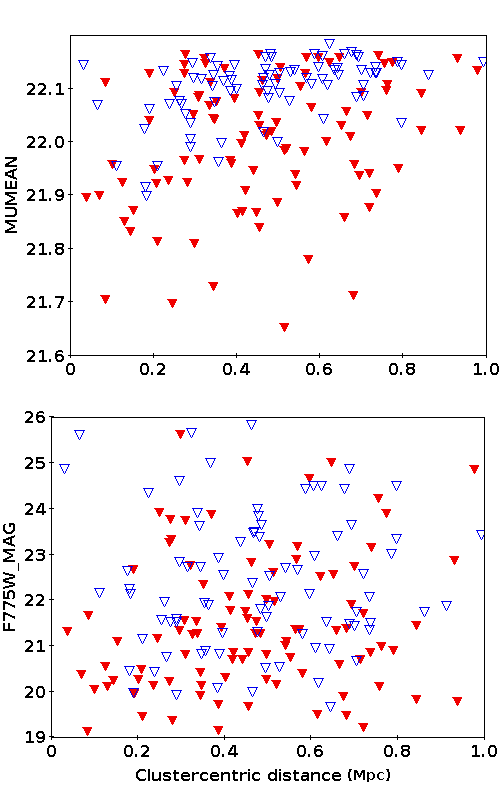}
    \caption{Relation between the surface brightness (top) and F775W magnitude (bottom) with clustercentric distance. For symbols description see Fig.~\ref{fig:SR}.)}
    \label{fig:D_two}
\end{figure}

\indent We also analysed the relation between the galaxy brightness, in terms of F775W magnitude and surface brightness (MUMEAN), and clustercentric distance, as shown in Fig.~\ref{fig:D_two}. For the two morphological types, Table ~\ref{tab:D} gives again the main statistics regarding the brightness.

Finally, we analysed the relation between the clustercentric distance (R) and morphological parameters measured in previous section. Fig.~\ref{fig:D_all} shows for the first time for ZwCl0024+1652 how the six morphological parameters vary with respect to the clustercentric distance in the case of cluster members classified as ET and LT. We also selected those sources classified as LTs in this work, and that taking into account previous studies of \cite{Parekh2015} and visual inspection seem to be mergers. We discussed all plots and statistics in Sec.~\ref{sec:Discussions}. 
\section{Results and Discussion}
\label{sec:Discussions}
\begin{figure*}
    \includegraphics[width=17cm]{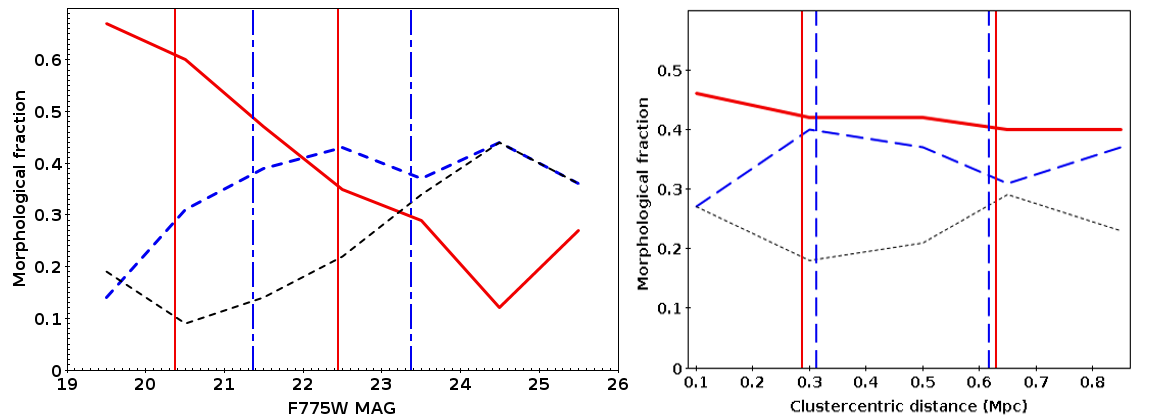}
    \caption{Morphological fractions as a function of F775W (left panel) with a binning size of 1\,MAG and as a function of clustercentric distance (right panel) with a binning size of 0.2\,Mpc along the x\,-\,axis.
    The thick red plot is for ET fraction, the blue dashed plot for the LT and the black thinner dashed plot for UD fraction. The thick red vertical 
    lines indicate the Q1 \& Q3 for ET fraction while the dashed blue vertical lines stand for the Q1 \& Q3 of the LT fraction.}. 
    \label{fig:morpho_frac}
\end{figure*}
\subsection{Morphological classes}
\label{sec:dis_morpho}
It was pointed out that the evolution of ET proportion is affected by redshift in addition to density and clustercentric distance 
(\citealt{Smith2005}; \citealt{Postman2005}; \citealt{Simard2009}). The BO effect (\citealt{BO1984}) verified with 
works for different redshift ranges, have been indifferently shown that LT proportion increases with redshift 
(e.g. \citealt{Fairley2002} for z\,$\sim\,$0.2 - 0.5, \citealt{deLucia2007} for z\,$\sim$\,0.4 - 0.8, \citealt{Barrena2012} for z\,$\sim$\,0.2 - 0.5 
and \citealt{Cast2014} for z\,$\sim$\,0.17 - 0.6). These studies show that the proportion of LTs at z\,$\sim$\,0.4 accounts about $\sim$35\,\% - 40\,\%. 
In our current work the fraction of LT galaxies is $\sim$36\,\%, which is in agreement with previous results. 

Moreover, according to \citet{Parekh2015} it was determined that galaxies are classified 
into most relaxed, relaxed and non-relaxed ones based on values of GINI coefficient where the non relaxed (peculiar/most disturbed) 
galaxies being characterized by GINI\,$<$\,0.4 criterion. It is also described for the most disturbed galaxies that GINI value is 
small because bright pixels are not compact while equally distributed in the given aperture radius. Accordingly eight non-relaxed 
galaxies (peculiar) were identified from LT class closer to the cluster core leaving the spiral population near the core to be very small.

For the overall galaxy population, \citet{Moran2007} determined that for 123 matching galaxies within the 1Mpc region, 65.6\,\% 
were ET while 34.1\,\% being LT. 
In the same work the morphology was determined for MS0451-0305 galaxy cluster at z\,$\sim$\,0.5 with 52\,\% and 48\,\% being ET and LT, respectively. The galaxy population in our work follows nearly the same trend as what has been on board for cluster studies 
confirming ET population is greater than the LT population (see \citealt{Postman2005} \& \citealt{Moran2007}). 
As shown in Sec~\ref{sec:class}, out of the 231 galaxies we have 42\,\% and 36\,\% galaxies classified as ET and LT, respectively.
\subsection{Morphology versus ELGs}
\label{sec:dis_morpho_ELG}
Using GLACE survey data, \citet{Sanchez2015} has presented a catalogue of 174 unique emission line galaxies (ELGs) 
in our cluster within 4\,Mpc clustercentric distance. Accordingly $\sim$\,37\,\% of the ELGs (64 galaxies) were shown to be AGN 
(broad line AGN (BLAGN) and narrow line AGN (NLAGN)) whereas $\sim$\,63\,\% being star forming (SF) galaxies (110 in number). 
Out of the 174 ELGs, 79 galaxies (52 SFs ($\sim$\,66\,\%) 
and 27 AGNs ($\sim$\,34\%)) were within the clustercentric distance of 1\,Mpc (region of our concern). 
Matching the GLACE result with ours we found 43 ($\sim$\,54.4\,\%) counterparts. Here 26 ELGs had no match in our catalogue may be because 
\citet{Sanchez2015} was working only on ELGs but this is not the case of our work. Out of the matching 43 sources, 26 ($\sim$\,60.5\,\%) are 
SF while the remaining 17 ($\sim$\,39.5\,\%) are AGN. 
Morphologically comparing the matching ELGs; 11 galaxies ($\sim$\,26\,\%) correspond to ET, 28 galaxies ($\sim$\,65\,\%) belong to LT and 
the remaining 4 galaxies ($\sim$\,9\,\%) correspond to UD class in our results. More specifically 18 SF galaxies are in LT class 
where 5 SF galaxies fall in ET while the remaining 3 SF galaxies belong to UD class. 
Similarly for AGN; LT contributes 10, ET contributes 6 while UD contributes only 1 AGN. 
This confirms that ELGs (SF as well as AGN) are mostly galaxies rather than ET galaxies.
\subsection{Morphological fractions}
\label{sec:dis_morpho_frac}
In Fig.~\ref{fig:morpho_frac} we compared the morphological fraction with both F775W magnitude and clustercentric distance.
To compare it with magnitude (left plot of Fig.~\ref{fig:morpho_frac}), morphological fraction is computed for each 1 magnitude 
bin in such a way that a particular class fraction is the ratio of the number of a given class galaxies to the total number of 
galaxies of all classes within the same bin. This for instance, is given for ET fraction in a given bin as:
\begin{equation}
 MAG\_ET_{frac}=\frac{number\,\,of\,\,ETs}{total\,\, number\, of\, galaxies},
 \label{eq:ET_frac}
\end{equation}
where number of ETs $\equiv$ ETs with magnitudes within the range of the bin; total number of galaxies $\equiv$ number of all 
galaxies (ET+LT+UD) with magnitudes in the same magnitude range.

Once it is computed for all the bins it is plotted 
against the centre of the bin. It can be seen that the fraction of ET galaxies decreases as a function of increasing magnitude 
while that of LT galaxies increases up to F775W\,$\sim$\,22.5, remaining nearly constant for fainter magnitudes.
The median F775W value for ET is determined to be 21.28 and that for LT is 22.16 while that for UD galaxies is 23.20. 
Hence we can see that the brightest galaxies are most likely to be resolved and classified into ET/LT whereas fainter galaxies could 
not easily be resolved, significant number of these galaxies are unlikely to be classified then left as UD. 
For magnitudes where F775W\,$>$\,24.5, the number of sources are very small that the statistics is very poor to draw a conclusion.

According to \citet{Fasano2012}, the fraction of ET galaxies is high near the centre of a nearby cluster while 
decreasing as a function of clustercentric distance. The fraction of LT galaxies on the other hand being smaller 
closer to the core while increasing as a function of 
clustercentric distance (see also \citealt{Zwicky1942}; \citealt{Dressler1980}; \citealt{Whitmore1993}; 
\citealt{PC2016}). Here to compare with clustercentric distance (right panel of Fig.~\ref{fig:morpho_frac}) we compute 
the morphological fraction in each 0.2\,Mpc bin for 0 to 0.6\,Mpc, 0.1\,Mpc bin for 0.6 to 0.7\,Mpc and 0.3\,Mpc bin for 0.7\,Mpc to 1\,Mpc 
in the same way as in eq.~\ref{eq:ET_frac}. Once it is computed for all the bins it is plotted against the centre of the bin. 
Results of our current work (see Fig.~\ref{fig:D_histo} and the right plot of Fig.~\ref{fig:morpho_frac}) confirm that 
closer to the core, the ET population fraction is higher than the LT fraction, but ET fraction is observed decreasing and LT 
fraction increasing until the clustercentric 
distance of $\sim$\,0.3\,Mpc. Whereas beyond $\sim$\,0.3\,Mpc fractions of both populations continue nearly flat 
in parallel up to a clustercentric distance; R\,$\sim$\,1\,Mpc. 
For clustercentric distances where R\,$>$\,0.7\,Mpc, the number of sources are very small that the statistics is very poor to conclude. 
We can see in general on Fig.~\ref{fig:morpho_frac} throughout the entire region that the fraction of ET galaxies is 
consistently higher than the LT fraction and more fraction of galaxies is classified into ET/LT near the core (lower UD fraction) than in 
far distances (where higher UD fraction) from the centre. Hence, our results are in a good agreement with previous results.

Moreover, out of the total number of 231 galaxies with in the cluster; 111 have spectroscopically confirmed redshifts while 120 have 
photometric redshifts. Different trends are observed in morphological fractions throughout the clustercentric distance. It can easily be seen 
that for galaxy population with spectroscopic redshifts, ET fraction is greater than the LT fraction throughout the region. While for galaxies 
with photometric redshifts, the LT fraction dominates throughout over the ET faraction. 
\subsection{Morphology-density relation}
\label{sec:dis_morpho_density}
An important point to be raised is morphology - density relation. As shown by \citet{Hoyle2012}, there is a trend of an increase in 
the population of ET galaxies 
towards the cluster centre accompanied by a strong morphology - density relation. Previous studies have already described a high - intermediate - low 
density regions in a cluster (see \citealt{Jee2005}; \citealt{Demarco2010}). 
Analysing a cluster at z\,$=$\,0.84, \citet{Nantais2013} determined that the cluster outskirts (intermediate to low density region) 
is characterised by higher LT while lower spiral with more peculiar (merging) galaxy population. Whereas  
a high density region (cluster core) with dominating ET population, few peculiar galaxies and almost devoid of spirals. 
It has been determined near the cores of clusters that the proportion of ET galaxies is $\sim$\,47\,\%, that is $\sim$\,2.8 times greater than 
the ET fraction in the field at the 
same intermediate redshift (see \citealt{Delgado2010} \& \citealt{Nantais2013}).
This is a relationship that holds also for low redshift rich clusters as determined for galaxies with redshift 
of z\,$\sim$\,0.1 - 0.2 by \citet{Fasano2000}.
In our case, we were working 
on the broad classes (ET and LT). While there is no clear classification into peculiar (merging) galaxies and these will be included in our 
classification either into LT or UD. In our work, the proportion of ET 
is recorded decreasing as going outwards (decreasing density) from the cluster core at least up to $\sim$\,0.7\,Mpc (see the right plot in 
Fig.~\ref{fig:morpho_frac}). As mentioned previously, after the R\,=\,0.7\,Mpc the number of sources decrease significantly in all 
three morphological groups, which affects the measured fractions. Moreover, LT population decreases approaching to the cluster 
core in agreement with existing results. 
\subsection{Relevance of morphological parameters}
\label{sec:dis_morpho_params}
In \citet{Parekh2015} while working on galaxy classification into relaxed versus dynamically disturbed system using the data of clusters at 
different redshifts from Chandra archive, they indicated GINI, M20 and Concentration as very promising parameters for identifying mergers. 
Accordingly, the criteria set for the most relaxed system is that GINI\,$>$\,0.65, M20\,$<$\,-2.0 and Concentration\,$>$\,1.55. For the most 
dynamically disturbed (non relaxed) system GINI\,$<$\,0.4, M20\,$>$\,-1.4 and Concentration\,$<$\,1 were set. 
Intermediate between the two extreme conditions is the mildly disturbed situation.
They identified that GINI is the most useful parameter in determining substructure because it does not depend on the exact position of 
the centre. Our classification was done with six morphological parameters (subsection ~\ref{sec:Parameters}) to classify the galaxies 
into ET and LT. Adapting the 
criteria from \citet{Parekh2015} for our work, GINI\,$<$\,0.4 gave us 12 galaxies classified into ET/LT (ET=1 (8.3\,\%), LT=11 (91.7\,\%)). 
These 11 LT galaxies are checked visually to be the most perturbed (non relaxed) galaxies, where M20\,$>$\,-1.4 for 9 galaxies out of 
the 11 LTs (with $\sim$\,82\,\% agreement). 
On the other hand corresponding to GINI\,$>$\,0.65; 58 galaxies were classified into ET/LT 
(ET=49 (85\,\%), LT=9 (15\,\%)) and are the most relaxed ones accordingly. Hence our result in this aspect is subject to 85\,\% agreement 
with previous works (amount of ETs). But in our work, M20\,$<$\,-2.0 is too small to be used while it is better to take M20 cut off for the 
most relaxed galaxies to be $\lesssim$\,-1.8 to establish an accuracy of at least 79\,\%. 
Moreover, since concentration parameters are defined as in subsection ~\ref{sec:Parameters}, from our results the cut off 
values of CABR\,$<$\,0.2 and CCON\,$<$\,7.0 can be used for the most perturbed galaxies with $\sim$\,91\,\% agreement while CABR\,$>$\,0.45 
and CCON\,$>$\,7.5 can segregate about $\sim$\,94\,\% of the most relaxed galaxy population. Therefore with this cut off limits,  
CCON and CABR parameters could also be equally important parameters as GINI and M20 for morphological classification of galaxies. 
\subsection{Morphological parameters vs. clustercentric distance}
\label{sec:dis_morpho_distance}
\begin{figure*}
    \includegraphics[width=17cm]{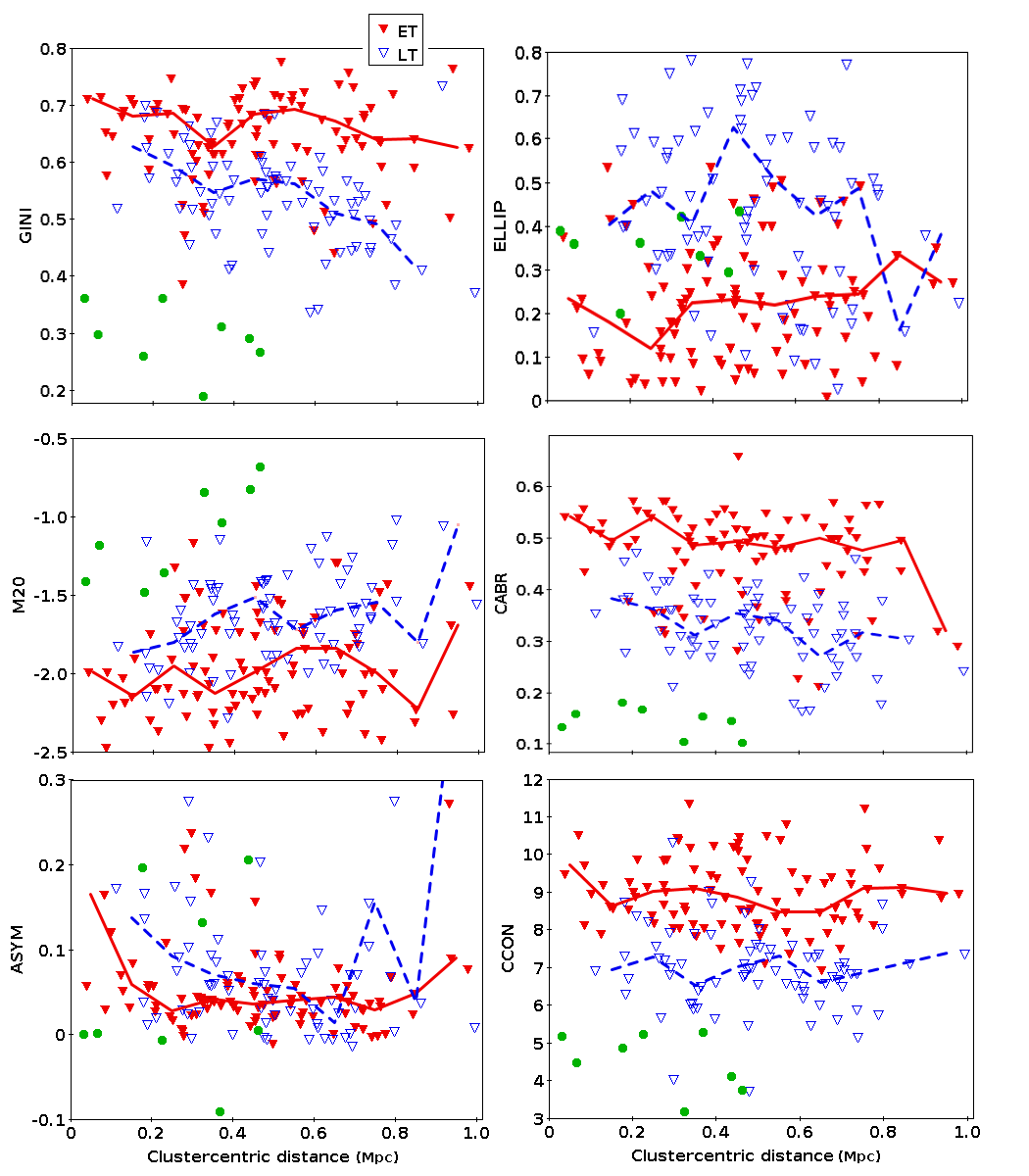}
    \caption{From top to bottom, and from left to right: Relation between the GINI, ellipticity, M20 moment of light, CABR concentration index, 
    asymmetry, and CCON concentration index and distance from the cluster centre. In all plots red solid triangles stand for ET, and blue open 
    triangles for LT galaxies. The disturbed (merging) galaxies selected from LT based on GINI\,$<$\,0.4 criteria are indicated with green dots 
    (solid circles). Median values of each parameter with clustercentric distance are shown with the red solid and blue dashed lines of ET and 
    LT galaxies, respectively.} 
    \label{fig:D_all}
\end{figure*}

In this work for the first time, we studied the properties of different morphological parameters in relation to the clustercentric distance (R). In Fig.~\ref{fig:D_all}
we showed how GINI, ELLIP, M20, CABR, ASYM and CCON change with R for ET and LT galaxies. In general, we do not find any clear trend in case 
of ASYM, CCON and ELLIP with R. In case of GINI, CABR and M20 a slight trend is observed of decreasing GINI and CABR showing and increasing 
M20 moment of light, suggesting that as going outwards from the cluster centre the light concentration decreases. However, much better statistics 
are needed to confirm this result. 
As shown in the right plot of Fig.~\ref{fig:morpho_frac} for ET class galaxies, the median and [Q1 - Q3] range of R being lower on average than LT, 
and this can be explained against each parameter. This is accompanied by higher values of GINI, CCON and CABR while lower values of ELLIP, M20 and 
values about zero for ASYM for ET galaxies. Whereas for LT galaxies, the median and [Q1 - Q3] range of R is slightly higher on average than ET galaxies.
This can be seen from the plot describing morphological fraction in Fig. ~\ref{fig:morpho_frac}. It can also be seen that GINI value slightly 
decreases as a function of increasing R for LT galaxies. Similar trend is observed from the plot of CABR 
versus R but very slow decrease for both classes in this case. For other parameters the values almost remain stagnant with 
R.
\section{Conclusions}
\label{sec:Summary}
In this work as part of a complete morphological study of the cluster ZwCl0024+1652 at an intermediate redshift z\,$\sim$\,0.4, we presented a broad classification of member galaxies with available 
redshifts within the clustercentric distance of 1\,Mpc using the HST/ACS image. 
We have classified galaxies up to the I - band magnitude of 26. By running galSVM code on a sample of 255 galaxies, 
6 morphological parameters were measured and classification was provided for 231 galaxies. Of these, 111 have spectroscopic and 120 photometric redshift measurements. From our classification and analysis we have drawn the following conclusions:
\begin{itemize}
  \item Out of all the 231 galaxies 97 ($\sim$\,42\,\%) were classified as ET, 83 ($\sim$\,36\,\%) as LT and 51 ($\sim$\,22\,\%) stayed unclassified. 
 If we take the well classified galaxies (180 in number); 97 ($\sim$\,54\,\%) were classified as ET whereas 83 ($\sim$\,46\,\%) fall into an LT class. 
  \item Comparing with the visual classification results in \citet{Moran2007} we have classified 53 galaxies matching with their previous 
  visual morphologies, 6 galaxies classified to different classes and 121 new sources which didn't have any reported morphological 
  classification within $\sim$\,1\,Mpc) radius are newly classified in our work. Therefore this work work gives the most complete and 
  largest morphological catalogue available up to now for our galaxy cluster. 
 \item Moreover, our comparison with the existing visual classification of \citet{Moran2007} is in a good 
 agreement of 81\,\%. Hence, applying galSVM for morphological classification can be taken as a reliable technique to be used even for a large sample.
 \item We have tested that ET and LT galaxies follow the expected distributions for different standard morphological diagrams, 
 colour - colour and colour - magnitude diagrams.
 \item The ET morphological fraction is higher near the cluster core decreasing outwards with LT fraction being lower at core increasing outwards. 
 Throughout the region of 1\,Mpc radius, the fraction of ET galaxies is consistently greater than the LT fraction for our cluster in the region 
 of our concern (R out to 1\,Mpc). Hence, the ET/LT fraction in the cluster is in agreement with previous studies.
 \item Morphological fractions in our galaxy cluster at z\,$\sim$\,0.4 evolves with magnitude in such a way that ET fraction dominates 
 in the brightest magnitude limit decreasing towards the fainter end while the LT fraction increases as magnitude goes fainter.  
 \item We compared our results with \cite{Sanchez2015} and found 43 ELG counterparts. 
 As a result out of these counterparts, 11 galaxies ($\sim$\,26\,\%) correspond to ET while 28 galaxies ($\sim$\,65\,\%) were found to belong to LT. 
with the remaining 4 galaxies ($\sim9\%$) stayed unclassified in our work. Moreover with the star forming ELGs; 18 SF galaxies are 
LT, 5 SF galaxies fall in ET and the remaining 3 SF galaxies belong to UD class. Similarly for AGN; LT contributes 10, ET contributes 6 
while UD contributes only 1 AGN. Hence, in general we deduce that ELGs are more of LT in morphology than ET.
 \item We have analysed the morphological parameters as a function of clustercentric distance out to 1\,Mpc for the first time. In general we do not find any clear trend, however better statistics would be valuable in future studies to revise the change of galaxy light concentration with R.
 \end{itemize}
 This work contributes significantly in the area of studies related to evolution of galaxies in clusters involving morphological classification, and provides the most complete morphological catalogue of ZwCl0024+1652. In our future studies within the GLACE survey, we are planning to compare morphological properties with metallicities, star formation, and AGN contribution using the tunable filters data. 
Finally, a complete morphological catalogue that resulted from our work can be accessed with an electronic version of this paper. The first seven rows 
and the descriptions for all the columns are presented as an appendix in this paper.  
\section*{Acknowledgements}
We acknowledge the anonymous referee for helpful comments and valuable suggestions that have significantly contributed to improve the paper. 
We also thank the Ethiopian Space Science and Technology Institute (ESSTI) under the Ethiopian Ministry of Innovation and Technology (MOIT) for all 
the financial and technical supports. ZBA specially acknowledges Joint ALMA Observatory (JAO); ESO\,-\,ALMA at Santiago, Chile for giving financial support 
and working space while working with M. Sanchez-Portal on the first phase of the paper. Moreover ZBA acknowledges Kotebe Metropolitan University for 
granting a study leave and giving material supports. MP and SBT acknowledge financial support from the Ethiopian Space Science and Technology Institute 
(ESSTI) under the Ethiopian Ministry of Innovation and Technology (MoIT). MP also acknowledges support from the Spanish MINECO under projects 
AYA2013-42227-P and AYA2016-76682-C3-1-P, and from the State Agency for Research of the Spanish MCIU through the ``Center of Excellence Severo Ochoa'' 
award for the Instituto de Astrof\'isica de Andaluc\'ia (SEV-2017-0709).
In this work, we made use of Virtual Observatory Tool for OPerations on Catalogues And Tables (TOPCAT) and IRAF. IRAF is distributed by the National Optical 
Astronomy Observatories, which are operated by the Association of Universities for Research in Astronomy, Inc., under cooperative agreement with the National
Science Foundation. We also used ACS/HST data based on observations made with the NASA/ESA HST, and obtained from the Hubble Legacy Archive, which is a
collaboration between the Space Telescope Science Institute (STScI/NASA), the Space Telescope European Coordinating Facility (ST-ECF/ESA) and the Canadian 
Astronomy Data centre (CADC/NRC/CSA).This work was supported by the Spanish Ministry of Economy and Competitiveness (MINECO) under the grants AYA2014\,-\,58861\,-\,C3\,-\,2
\,-\,P, AYA2014\,-\,58861\,-\,C3\,-\,3\,-\,P, AYA2017\,-\,88007\,-\,C3\,-\,1\,-\,P and AYA2017\,-\,88007\,-\,C3\,-\,2\,-P.








\newpage
\appendix

\section{The comprehensive morphological catalogue of galaxies in ZwCl0024+1652 cluster at $z\sim0.4$}

A complete morphological catalogue of 231 sources (111 with spectroscopically confirmed redshifts while 120 with photometric redshifts) is presented 
in this work. This is the most comprehensive catalogue containing the morphological classes of galaxies in ZwCl0024+1652 cluster; classified with galSVM technique. The entire catalogue comprises of a table of 231 rows standing 
for sources (galaxies) and 34 columns with respective parameter for each row. In this catalogue, the morphological class is identified for 
180 galaxies but the remaining 51 galaxies are marked undecided in terms of the morphological class. In addition to parameters measured in this work, 
SExtractor measured photometric data and results from previous works \citealt{Treu2003}; \citealt{Moran2005}; \citealt{Moran2007} 
and \citealt{Sanchez2015}) is also included in this catalogue. The followings are descriptions of each 
column of the entire catalogue.\\
\textbf{Column 1} ... Source index (galaxy number);\\
\textbf{Column 2} ... HST identification number of the galaxy (99.0 if not available);\\
\textbf{Column 1} ... Source index (galaxy number);\\
\textbf{Column 2} ... HST identification number of the galaxy (99.0 if not available);\\
\textbf{Column 3} ... Right Ascension in decimal degrees (J2000);\\
\textbf{Column 4} ... Declination in decimal degrees (J2000);\\
\textbf{Column 5} ... Ellipticity of the galaxy measured by SExtractor;\\
\textbf{Column 6} ... MUMEAN value for the galaxy measured by SExtractor;\\
\textbf{Column 7} ... Asymmetry index measured by galSVM (described in subsection ~\ref{sec:Parameters});\\
\textbf{Column 8} ... Abraham concentration index measured by galSVM (described in subsection ~\ref{sec:Parameters});\\
\textbf{Column 9} ... GINI coefficient measured by galSVM (described in subsection ~\ref{sec:Parameters});\\
\textbf{Column 10} ... M20 moment of light index measured by galSVM (described in subsection ~\ref{sec:Parameters});\\
\textbf{Column 11} ... Bershady-Concelice concentration index measured by galSVM (described in subsection ~\ref{sec:Parameters});\\
\textbf{Column 12} ... MAG\_AUTO (F775W) (we measured it by SExtractor for each source); \\
\textbf{Column 13} ... Uncertainity in MAG\_AUTO (F775W) measured by sextractor for each source;\\
\textbf{Column 14} ... Redshift values for each galaxy from the public data of ZwCl0024+1652 master catalogue generated by a team working 
			      on a "A Wide Field Survey of Two $z=0.5$ Galaxy Clusters"  (see \citealt{Treu2003} and \citealt{Moran2005}).\\
\textbf{Column 15}... Zsource (=6 from DEIMOS, 1-5: other spectroscopic sources (see \citealt{Moran2007}) 7=secure photo-z (see \citealt{Smith2005}, 8=fairly unreliable photo-z (fewer bands,fainter);\\
\textbf{Column 16} ... Distance of the galaxy from the centre of the cluster (clustercentric distance) measured in Mpc;\\
\textbf{Column 17} ... The final probability computed taking the uncertainity (error) into account (used for 
			      morphologically classifying the galaxies in current work);\\
\textbf{Column 18} ... The final morphological class (Early\_Type (ET), Late\_Type (LT) or Undecided (UD)) based on the final probability value\\
\textbf{Column 19} ... Visual morphology as given in \citet{Moran2007}; $=-99.9$ if not available;\\
\textbf{Column 20} ... Emission Line Galaxy (ELG) type adapted from \citet{Sanchez2015}; $=-99.9$ if not available;\\
\textbf{Columns 21 - Column 34} ... SExtractor measured photometric data (MAG\_AUTOs and errors) for the galaxies taken 
					   from public data of ZwCl0024+1652 master catalogue (see \citealt{Treu2003} 
					   and \citealt{Moran2005}); $=99.$ if not available,$=-99.9$ if all values not measured.\\ 
 Part of the catalogue (the column values for the first seven sources) is presented in Table~\ref{tab:Catalogue}. 
 For sample illustration, the first 7 rows of the electronic version of the catalogue with values of the columns (all the parameters) 
 are presented here.
	\begin{table*}
	\center
	\caption{Part of the Morphological Catalogue of galaxies in ZwCl0024+1652 Cluster (Full catalogue is available online)}
	\label{tab:Catalogue}
	\begin{tabular}{lllllllll} 
		\hline\hline
		NUMBER & HST\_ID & RA (deg) & DEC (deg) & ELLIP & MUMEAN \\ 
		ASYM & CABR & GINI & M20 & CCON & F775W\_Mag \\ 
		F775W\_ERR & Redshift & Zsource & R(Mpc) & PROBA\_FINAL & GALAXY\_CLASS \\
		Visual\_Morpho & ELG\_Type & B\_AUTO & V\_AUTO & R\_AUTO & I\_AUTO \\
		J\_AUTO & K\_AUTO & F814W\_AUTO & B\_ERR & V\_ERR & R\_ERR \\
		I\_ERR & J\_ERR & K\_ERR & F814W\_ERR \\ 
		\hline\hline
		\textbf{1} & 80.0 & 6.62029 & 17.13273 & .2272 & 22.1522 \\   
		.0087 & .2443 & .3740 & -1.5537 & 7.3720 & 23.4480 \\
		.3906 & .3810 & 8 & .9763 & .3049 & LT \\
		-99.9 & -99.9 & 25.7701 & 24.9023 & 23.7896 & 23.0985 \\
		21.5017 & 19.9099 & 22.7271 & .1837 & .1923 & .0870 \\
		.0974 & 99. & 99. & .0268 \\
		\hline
		\textbf{2} & 99.0 & 6.65543 & 17.13758 & .5847 & 22.1088 \\
		.5657 & .3714 & .5321 & -1.8222 & 6.9793 & 20.7002 \\
		.1101 & .3940 & 1 & .6997 & .1017 & LT \\
		-99.9 & -99.9 & 23.4553 & 22.4472 & 21.1820 & 20.3412 \\
		18.9132 & 17.3511 & 99.0 & .0251 & .0224 & .0090 \\
		.0086 & .0551 & .0561 & 99. \\ 
		\hline
		\textbf{3} & 116.0 & 6.63273 & 17.13629 & .2521 & 22.1632 \\
		-.0019 & .3118 & .4927 & -1.5734 & 8.3250 & 23.1702 \\
		.3437 & .3660 & 8 & .7356 & .7600 & ET \\
		-99.9 & -99.9 & 25.2051 & 24.4749 & 23.4430 & 22.9775 \\
		21.4379 & 19.8460 & 22.7281 & .1164 & .1378 & .0672 \\
		.0924 & 99. & 99. & .0120 \\           
		\hline
		\textbf{4} & 126.0 & 6.64078 & 17.13619 & .2470 & 22.1618 \\
		.0642 & .2593 & .4668 & -1.4197 & 7.7667 & 24.5245 \\
		.6413 & .3780 & 8 & .6845 & .5944 & UD \\
		-99.9 & -99.9 & 26.9703 & 26.0467 & 25.7980 & 23.5168 \\
		21.5139 & 19.9220 & 23.8584 & 99. & 99. & 99. \\
		.1415 & 99. & 99. &  .0234 \\           
		\hline
		\textbf{5} & 9.0 & 6.66708 & 17.13923 & .3327 & 22.0930 \\
		.0482 & .4381 & .5917 & -2.2275 & 8.9588 & 21.4544 \\
		.1559 & .3999 & 6 & .8261 & .7411 & ET \\
		S0 & NLAGN & 24.1371 & 22.8917 & 21.7965 & 21.1522 \\
		20.1556 & 18.8312 & 20.9172 & .0398 & .0291 & .0136 \\
		.0155 & .1482 & .1887 & .0079 \\           
		\hline
		\textbf{6} & 99.0 & 6.65664 & 17.13732 & .4107 & 22.1440 \\
		-.0049 & .3036 & .5818 & -1.3960 & 10.8122 & 24.0350 \\
		.5118 & .4470 & 8 & .7183 & .5125 & UD \\
		-99.9 & -99.9 & 25.2620 & 25.2038 & 24.0649 & 23.5652 \\
		21.4067 & 19.8148 & 99.0 & .1260 & .2766 & .1221 \\
		.1632 & 99. & 99. & 99. \\             
  		\hline
		\textbf{7} & 58.0 & 6.64736 & 17.13850 & .0849 & 22.1506 \\
		-.1024 & .3391 & .5138 & -1.8383 & 7.7022 & 22.5379 \\
		.2568 & .3840 & 7 & .6226 & .6945 & ET \\
		-99.9 & -99.9 & 24.9017 & 24.1310 & 23.2602 & 22.2188 \\   
		21.5017 & 19.9099 & 22.0129 & .0836 & .0950 & .0537 \\
		.0435 & 99. & 99. & .0074 \\
		\hline
	\end{tabular}
	\end{table*}


\bsp	
\label{lastpage}
\end{document}